\def\mode{0} % print

\if 1\mode
    \documentclass[preprint,superscriptaddress,floatfix]{revtex4-2}
    \usepackage{lineno}
    \linenumbers
    \usepackage{natbib}
\else
    \documentclass[twocolumn,superscriptaddress]{revtex4-2}
\fi

\usepackage{graphicx}% Include figure files
\usepackage{dcolumn}% Align table columns on decimal point
\usepackage{bm}% bold math
\usepackage{float}
\usepackage{siunitx}
\usepackage{ amsmath,amssymb}
\usepackage[
	pdftex,
	colorlinks=true,
  	urlcolor=blue,
  	linkcolor=blue,
  	citecolor=blue
]{hyperref}

\newcommand{\Er}{\ensuremath{E_{\mathrm{R}}}}

\begin{document}

\title{Dipolar quantum solids emerging in a Hubbard quantum simulator}

\author{Lin Su}
\affiliation{Department of Physics, Harvard University, Cambridge, Massachusetts 02138, USA}
\author{Alexander Douglas}
\affiliation{Department of Physics, Harvard University, Cambridge, Massachusetts 02138, USA}
\author{Michal Szurek}
\affiliation{Department of Physics, Harvard University, Cambridge, Massachusetts 02138, USA}

\author{Robin Groth}
\affiliation{Department of Physics, Harvard University, Cambridge, Massachusetts 02138, USA}
\author{S. Furkan Ozturk}
\affiliation{Department of Physics, Harvard University, Cambridge, Massachusetts 02138, USA}
\author{Aaron Krahn}
\affiliation{Department of Physics, Harvard University, Cambridge, Massachusetts 02138, USA}
\author{Anne H. H\'{e}bert}
\affiliation{Department of Physics, Harvard University, Cambridge, Massachusetts 02138, USA}
\author{Gregory A. Phelps}
\affiliation{Department of Physics, Harvard University, Cambridge, Massachusetts 02138, USA}
\author{Sepehr Ebadi}
\affiliation{Department of Physics, Harvard University, Cambridge, Massachusetts 02138, USA}

\author{Susannah Dickerson}
\affiliation{Department of Physics, Harvard University, Cambridge, Massachusetts 02138, USA}

\author{Francesca Ferlaino}
\affiliation{Institut für Experimentalphysik, Universität Innsbruck,Technikerstraße 25, 6020 Innsbruck, Austria}
\affiliation{Institut für Quantenoptik und Quanteninformation, Österreichische Akademie der Wissenschaften, 6020 Innsbruck, Austria}
\author{Ognjen Markovi\'{c}}
\affiliation{Department of Physics, Harvard University, Cambridge, Massachusetts 02138, USA}
\author{Markus Greiner}
\if 0\mode
    \email{greiner@physics.harvard.edu}
\fi
\affiliation{Department of Physics, Harvard University, Cambridge, Massachusetts 02138, USA}

\date{\today}
\begin{abstract}

In quantum mechanical many-body systems, long-range and anisotropic interactions promote rich spatial structure and can lead to quantum frustration, giving rise to a wealth of complex, strongly correlated quantum phases \cite{Defenu2021}. Long-range interactions play an important role in nature; however, quantum simulations of lattice systems have largely not been able to realize such interactions. A wide range of efforts are underway to explore long-range interacting lattice systems using polar molecules \cite{Kaufman2021, Bohn2017, Chae2022, Bao2022,Holland2022, Christakis2023, Schindewolf2022, Bigagli2023, Rosenberg2022, Li2023}, Rydberg atoms \cite{Kaufman2021, Browaeys2020, Chen2023, Ebadi2021, Guardado-Sanchez2021}, optical cavities \cite{Mivehvar2021, Landig2016, Li2021, Guo2021}, and magnetic atoms \cite{Chomaz2022, Baier2016, Patscheider2020, Lepoutre2019}. Here, we realize novel quantum phases in a strongly correlated lattice system with long-range dipolar interactions using ultracold magnetic erbium atoms. As we tune the dipolar interaction to be the dominant energy scale in our system, we observe quantum phase transitions from a superfluid into dipolar quantum solids, which we directly detect using quantum gas microscopy with accordion lattices. Controlling the interaction anisotropy by orienting the dipoles enables us to realize a variety of stripe ordered states. Furthermore, by transitioning non-adiabatically through the strongly correlated regime, we  observe the emergence of a range of metastable stripe-ordered states. This work demonstrates that novel strongly correlated quantum phases can be realized using long-range dipolar interaction in optical lattices, opening the door to quantum simulations of a wide range of lattice models with long-range and anisotropic interactions.

\end{abstract}

\maketitle

\textit{Introduction.} Quantum simulations \cite{Altman2021} with ultracold atoms in optical lattices enable the exploration of strongly correlated quantum matter described by the Hubbard model \cite{Gross2017} and are reaching regimes that are extremely challenging to access numerically \cite{Bohrdt2021}. Quantum simulations of the Hubbard model, however, have so far largely been limited to local on-site interactions, and it has been a long-standing goal to realize simulations with strong long-range interactions. This would allow for the quantum simulation of models that more accurately describe realistic quantum materials like transition metal dichalcogenides, which typically experience finite off-site Coulomb repulsion \cite{Mak2022}. Tunable anisotropic long-range interactions would furthermore open the door to modeling quantum materials such as spin ice \cite{Castelnovo2012}, anisotropic materials \cite{Li2019}, or twisted bilayer materials \cite{Andrei2020}, as well as a wide range of models beyond current quantum material realizations. Long-range interactions naturally promote spatial structure, leading to the emergence of solid phases. In stark contrast to short-ranged Hubbard models, long-range interactions can lead to frustration in otherwise non-frustrated geometries, hosting supersolids \cite{Boninsegni2012, Wu2020, Capogrosso-Sansone2010, Bruder1993, Batrouni1995}, spin liquids \cite{Yao2018}, and fractionalization \cite{Mao2022, Prem2018}. Generally, the fundamental question arises: how do long-range interactions compete with kinetic energy and on-site interaction to give rise to novel quantum phases of matter?

Developing approaches to realize long-range interacting systems and address such a question is an exceptionally active field of research. Significant progress has been made to create controllable systems of cold polar molecules \cite{Kaufman2021, Bohn2017, Chae2022, Bao2022, Holland2022, Christakis2023, Schindewolf2022, Bigagli2023}, with ongoing efforts towards itinerant lattice gases \cite{Li2023, Rosenberg2022}. Rydberg interactions have enabled the quantum simulations of programmable Ising and XY spin models \cite{Kaufman2021, Browaeys2020, Chen2023, Ebadi2021}. However, using Rydberg dressing to realize itinerant models \cite{Guardado-Sanchez2021} presents challenges due to Rydberg decay. Dynamic light fields in optical cavities enable the study of infinite-range \cite{Landig2016,Li2021, Mivehvar2021} and finite-range \cite{Guo2021} interacting atom systems, with dissipation being the main hurdle for reaching strongly correlated lattice physics. Furthermore, recent experiments in condensed matter systems have simulated long-range interacting Hubbard systems with bulk measurements \cite{Lagoin2022, Li2022, Wang2022, Hensgens2017, Kennes2021}; yet, it is challenging to perform site-resolved studies in these simulators.

\begin{figure*}
    \centering
    \includegraphics[width=\textwidth]{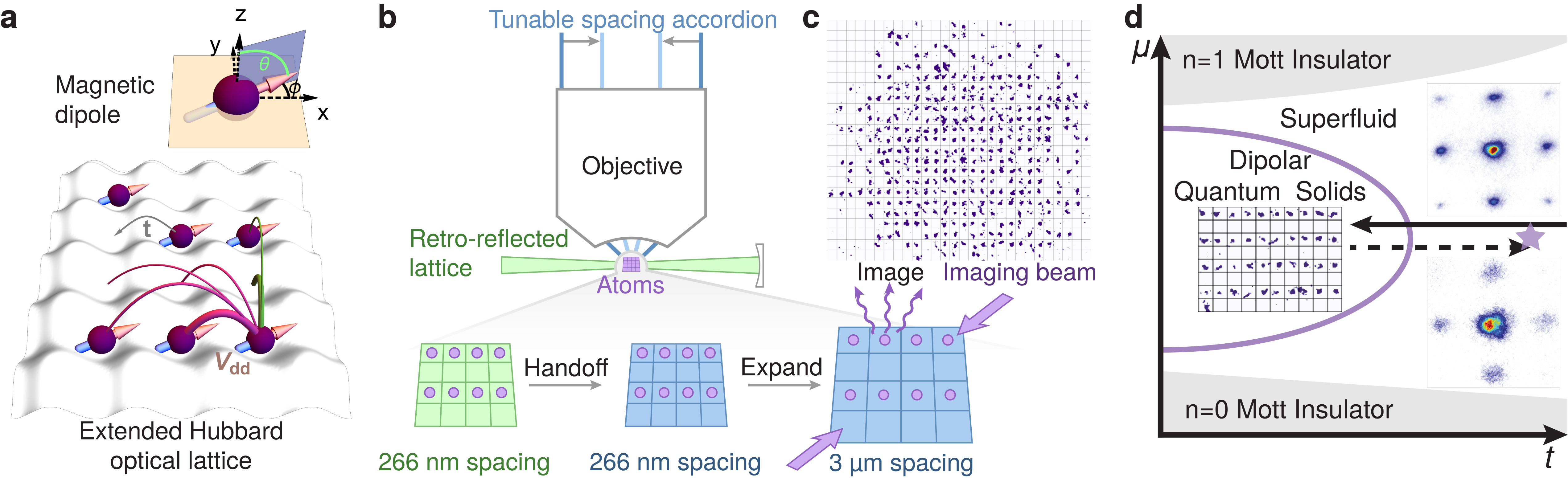}
    \caption{\textbf{Experimental setup.} Magnetic erbium atoms in an optical lattice realize an extended Hubbard model with anisotropic long-range interaction, enabling the realization of novel strongly correlated quantum phases. \textbf{a}, Orienting the magnetic dipoles using an external magnetic field allows us to widely tune the anisotropy of the interaction, with $\theta$ being the polar angle, $\phi$ the azimuthal angle,  the pink (green) lines denoting attractive (repulsive) dipole-dipole interaction $V_{dd}$, and the gray arrow denoting tunneling $t$.  \textbf{b}, We use a small-spacing lattice to maximize dipolar interaction strength. Quantum gas microscopy is realized by handing off atoms from the small-spacing retro-reflected lattices (green) to tunable spacing accordion lattices (blue). \textbf{c}, After expanding the accordion lattice, fast fluorescence imaging results in site-resolved single-shot images. \textbf{d}, We show a schematic of the extended Bose-Hubbard phase diagram with dipolar interactions.  The solid arrow indicates the ramp we perform to induce a quantum phase transitions from a strongly correlated superfluid to a half-filling dipolar quantum solid. As insets, we show an exemplary single-shot image of a solid (same as Fig. 2IIc) and time-of-flight images taken at the star location before the ramp (top) and after ramping back to the superfluid phase along the dashed arrow (bottom). The persistence of the superfluid matter wave peaks after ramping back from the solid qualitatively demonstrates the adiabaticity of the ramp.
    }
    \label{fig:figure1}
\end{figure*}

Magnetic atoms provide an attractive alternative to the systems mentioned above, as they are relatively simple to control while interacting through the long-range magnetic dipole-dipole interaction \cite{Lahaye2009}. In bulk systems, the atoms collectively order in the presence of magnetic interaction, leading to intriguing effects such as macroscopic dipolar droplets forming a supersolid phase \cite{Chomaz2022}. In order to realize strongly correlated physics, however, the long-range interaction energy between pairs of individual atoms needs to become large compared to the kinetic energy \cite{Dutta2015}. It was shown that a small-spacing optical lattice of magnetic atoms leads to a Hubbard system with dipolar interactions large enough to cause a significant shift of the superfluid to Mott insulator transition \cite{Baier2016}. However, whether the inter-site dipolar interaction can experimentally be made the dominant energy scale larger than kinetic energy and temperature remained an open question. Here, we present an affirmative answer to this question by observing quantum phase transitions from a strongly correlated superfluid to dipolar quantum solids, i.e. phases of matter that show spontaneous periodic density modulation in the presence of quantum fluctuations due to finite tunneling. The long-range and anisotropic interaction is expected to give rise to a rich set of states with fractional fillings \cite{Menotti2007, Capogrosso-Sansone2010, Zhang2015, Wu2020, Goral2002, Danshita2009}. In this work, we probe such states at half-filling.

We realize the 2D extended Bose-Hubbard model with anisotropic long-range dipolar interactions by employing magnetic erbium atoms in an optical lattice (Fig. 1a). In this work, we choose the atomic on-site Hubbard repulsion to be much larger than other energy scales. This realizes an extended Hubbard model in the hard-core boson limit in which each lattice site is either populated by 0 or 1 atom (see Methods). The Hubbard Hamiltonian of our system is

\begin{equation}
\label{eq:Hamiltonian}
H=-t\sum_{\langle i,j \rangle}(\hat{a}_i^\dag \hat{a}_j+\mathrm{h.c.})-\sum_i\mu_i\hat{n}_i+\sum_{i<j}V_{i,j}\hat{n}_i\hat{n}_j.
\end{equation}

Here, $\hat{a}_i^\dag\hat{a}_j$ describes the tunneling of hard-core bosons between nearest-neighbor sites with amplitude $t$, $\hat{n}_i$ is the number operator on site $i$, and $\mu_i$ is the chemical potential on site $i$. The dipolar interaction between lattice sites $i$ and $j$ with distance $\mathbf{d}_{i,j}=(d_x,d_y)$ sites is $V_{i,j}=V_0\frac{1-3((d_x/d)\sin\theta\cos\phi+(d_y/d)\sin\theta\sin\phi)^2}{d^3}$, where $(\theta,\phi)$ is the dipole polar and azimuthal angle (Fig. 1a), $d=|\mathbf{d}_{i,j}|$, and $V_0$ is the nearest-neighbor repulsive interaction energy when the dipole is oriented out of the 2D plane ($\theta=\qty{0}{\degree}$).

The major challenge for the observation of dipolar quantum solids using magnetic atoms is posed by the relatively weak magnetic interaction between atoms compared to the typical energy scales in quantum simulators. We reach a nearest-neighbor interaction energy of $V_0\approx h \times \qty{30}{\Hz}$,  where $h$ is Planck's constant, by using a small-spacing optical lattice (green beams in Fig. 1b) with $a=\qty{266}{\nm}$ lattice spacing \cite{Baier2016}. In order to resolve individual sites in this lattice, we developed a novel quantum gas microscopy technique using a tunable spacing two-dimensional accordion lattice \cite{Li2008} projected through a high numerical aperture in-vacuum objective (blue beams in Fig. 1b). For imaging, we transfer atoms from the small-spacing lattice to the accordion lattice and then expand the spacing of the accordion lattice. Finally, we image with single-site resolution (see Methods); a snapshot of a Mott insulator is shown in Fig. 1c.

\begin{figure*}
    \centering
    \includegraphics[width=\textwidth]{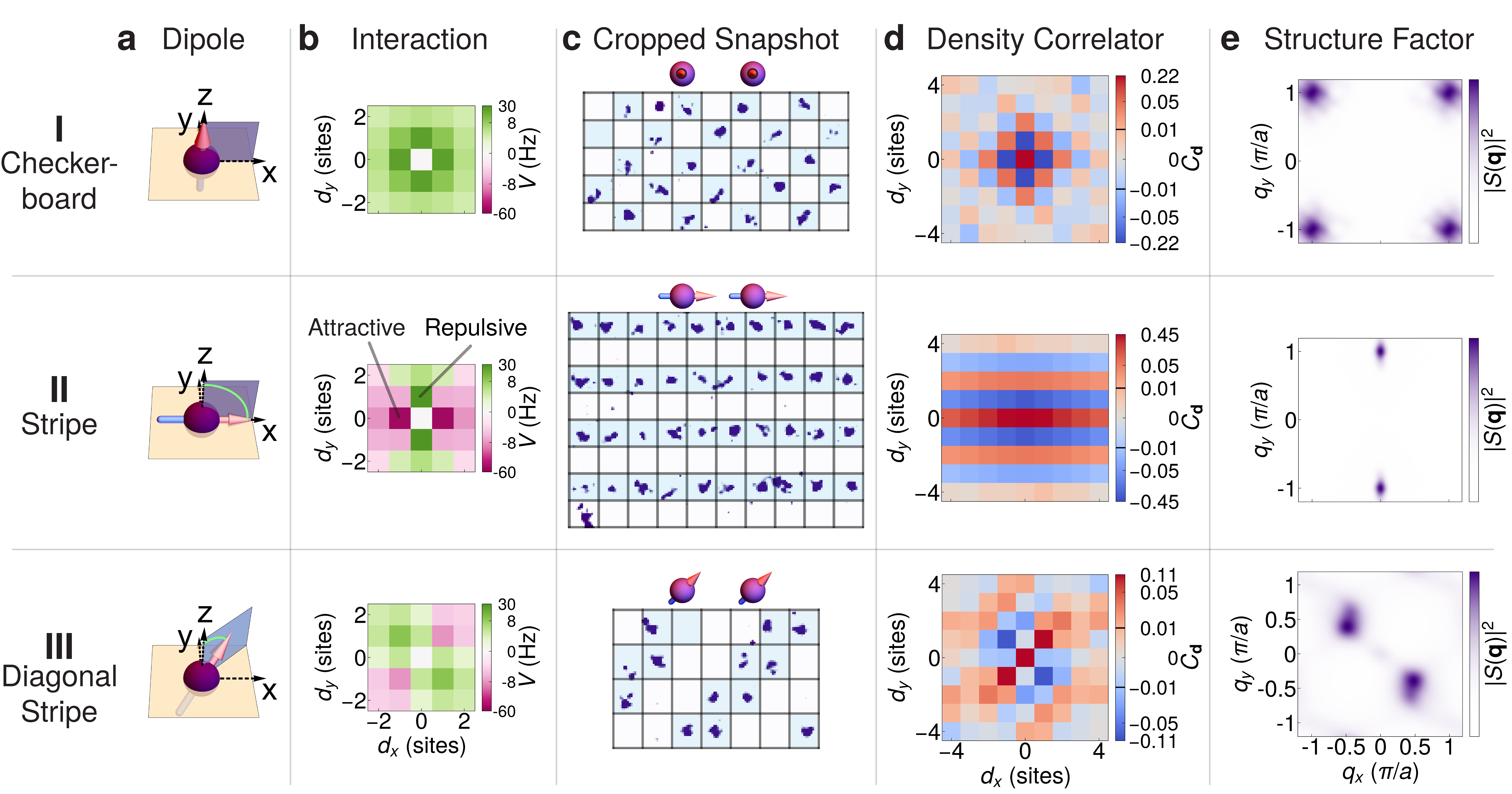}
    \caption{
    \textbf{Dipolar quantum solids}. As the dipolar interaction $V_0$ becomes dominant over tunneling $t$, dipolar quantum solids emerge. (\textbf{I}) Dipoles pointing in $z$-direction (a) give rise to isotropic $1/d^3$ repulsion (b). An exemplary single shot image (c) shows the checkerboard solid arising at half-filling (blue shading to guide the eye). The emerging order can be analyzed quantitatively by calculating the connected density-density correlator $C_d$ averaged over experimental realizations (d), with a linear color bar between  $-0.01$ to $0.01$ and logarithmic elsewhere. A Fourier transform of $C_d$ yields the structure factor $S(\mathbf{q})$, whose magnitude squared is plotted in (e) with a linear color bar starting from zero. (\textbf{II}) With the dipole aligned in the $x$-direction, the interaction becomes highly anisotropic, and the atoms form a stripe solid. (\textbf{III}) For diagonally aligned dipoles (here $\theta=$\qty{50}{\degree} and $\phi=$\qty{45}{\degree}) we observe diagonally ordered states. Remarkably, the transition into the diagonal stripe solid is entirely driven by beyond nearest-neighbor interaction terms.}
    \label{fig:figure2}
\end{figure*}

The dipolar interaction strength is nearly four orders of magnitude smaller than the optical lattice depth; hence, it is essential to  minimize lattice disorder causing the disorder in chemical potential $\mu_i$. We reduce the potential disorder to within a fraction of $V_0$ by minimizing the scattered light intensity in the atom plane and ensuring the residual disorder length scale is large. This is done by making sure that all optical surfaces on which lattice beams are incident are at least a few Rayleigh ranges away from the atomic plane \cite{Phelps2019}. This necessitates a custom objective with a central bore to minimize scattering of the vertical lattice beam (see Methods).

For the experiment, we start with a Bose-Einstein condensate of $^{168}$Er created in less than one second via narrow line laser cooling \cite{Phelps2020}. We then load the atoms into a single layer of a retro-reflected vertical lattice (see Methods). We robustly load sufficient atoms into the lattice to reach around 100 atoms on 200 sites in the central region of interest. The atomic dipole direction is set by aligning the dipoles to a tunable external magnetic field that is static during the lattice ramp. The small-spacing lattice power is ramped adiabatically to reach the target lattice depth following the solid arrow in Fig. 1d, such that we achieve $t/V_0 \approx 0.1$. Remarkably, we are able to remain adiabatic down to low tunneling strengths, corresponding to adiabatic cooling \cite{Pollet2008} to temperatures ($T$) of only several hundred picokelvin ($k_BT/V_0 \approx 0.5$). For imaging, we rapidly increase the lattice depth to turn off the tunneling in the small-spacing lattice and project the state of the system into the occupation basis. We then hand off the atoms to the accordion lattice and expand the lattice spacing to \qty{3}{\um}. The lattice dynamics are kept frozen when expanding the accordion lattice spacing such that the imaged lattice occupation faithfully represents the state of the atoms in the small-spacing lattice (see Methods). We then perform fluorescence imaging by exposing the atoms to resonant highly-saturated light for \qty{8}{\us} \cite{Bergschneider2018}. A few hundred fluorescence photons per atom are scattered and collected by the objective (Fig. 1b). About 50 photons per atom are detected by the electron-multiplying CCD camera, forming an image from which we extract atom occupation number per lattice site. The total experiment cycle time is \qty{2.5}{s}.

\textit{Dipolar quantum solids.}  We observe the atoms self-organizing into different dipolar quantum solids when the dipole-dipole interactions become the dominant energy scale in our system (Fig. 2). These periodically ordered states arise after quantum phase transitions from a strongly correlated superfluid into a dipolar quantum solid, with possible intermediate phases that will be the focus of future work. We transition into the solid by adiabatically reducing the kinetic energy of the atoms in the lattice via a linear ramp of lattice depth. The periodic ordering, which is distinct from the periodicity of the lattice, is a signature that the underlying lattice symmetry is broken. By tuning the orientation of the dipoles via the external magnetic field, we realize a wide variety of isotropic and anisotropic interactions leading to phases with different periodic ordering depending on the dipole orientation (Fig. 2, panels I, II, and III). We analyze the quantum solids directly from single-shot images or by calculating the connected density-density correlation

\begin{equation}
\label{eq:correlation}
C_\mathbf{d}=\frac{4}{N_{\mathbf{d}}}\sum_{\mathbf{d}=\mathbf{d}_{i,j}}(\langle \hat{n}_i\hat{n}_j\rangle-\langle \hat{n}_i\rangle\langle \hat{n}_j\rangle),
\end{equation}

where the sum runs over $N_{\mathbf{d}}$ pairs of lattice sites at distance vector $\mathbf{d}$ in the analysis region and the average runs over hundreds of experimental runs \cite{Mazurenko2017}. Such a correlation shows strong periodic patterns depending on the specific quantum solid (Fig. 2d). The Fourier transform of the correlation (the structure factor)

\begin{equation}
\label{eq:structure_factor}
S(\mathbf{q})\propto\sum_{\mathbf{d}}e^{i\mathbf{q}\cdot\mathbf{d}}C_\mathbf{d},
\end{equation}

exhibits peaks at positions set by the periodicity of the quantum solid state (Fig. 2e).

\textit{Checkerboard solid.} First, we tune the dipole orientation to explore isotropic long-range repulsion. When the atomic dipoles point perpendicular to the plane, i.e. $\theta=\qty{0}{\degree}$ (Fig. 2Ia), the atoms isotropically repel each other with a strength that decays as $1/d^3$, where $d$ is the distance between two lattice sites (Fig. 2Ib). At half-filling and weak tunneling, the atoms arrange themselves in the energetically favored checkerboard pattern. In Fig. 2Ic, we show a checkerboard in a cropped region of an example single-shot image. The periodic checkerboard structure in the observed density correlation (Fig. 2Id) and the structure factor exhibiting $(\pm\pi/a,\pm\pi/a)$ peaks (Fig. 2Ie) is a hallmark of lattice symmetry breaking in a solid \cite{Girvin2019}.

\textit{Stripe solid.} Next, we maximize the interaction anisotropy between the two lattice axes by orienting the atomic dipoles in the 2D atom plane along one of the lattice directions, i.e. $(\theta, \phi)=(\qty{90}{\degree}, \qty{0}{\degree})$ (Fig. 2IIa). In single-shot images, we observe long chains of atoms aligned with the atomic dipole direction, with an example cropped image shown in Fig. 2IIc. The phase of the chains changes from shot to shot (see Methods), showing the characteristics of spontaneous symmetry-breaking. The extracted connected density-density correlation shows a periodic stripe pattern (Fig. 2IId) and the density structure factor (Fig. 2IIe) clearly exhibits $(0,\pm\pi/a)$ peaks.

Furthermore, we probe the adiabaticity of the transition from the superfluid to dipolar quantum solids by measuring the superfluid coherence peaks before and after the lattice ramp. The superfluid fraction can be qualitatively estimated by observing coherence peaks in time-of-flight images, after releasing atoms from the lattice \cite{Greiner2002}. Fig. 1d inset, above the solid arrow, is a time-of-flight image before ramping into the stripe solid and shows the superfluid coherence peaks, demonstrating the quantum coherence of our initial state. To demonstrate the ramp into the ordered state is nearly adiabatic, we perform a closed path in parameter space and ramp back into the superfluid phase following the dashed arrow shown in Fig. 1d. We qualitatively check the return fidelity by taking a time-of-flight image shown below the dashed arrow. The appearance of interference peaks indicates that ramp to the solid is close to adiabatic.

\textit{Diagonal stripe solid.} To probe the long-range nature of the dipole-dipole interaction, we align the atomic dipole out of the lattice plane along the $(\theta, \phi)=(\qty{50}{\degree}, \qty{45}{\degree})$ direction. If our interactions only included nearest-neighbor interactions, this dipole orientation with repulsive nearest-neighbor interactions (Fig. 2IIIb) would be similar to the first case with $\theta=\qty{0}{\degree}$ (Fig. 2Ib), where we observe the checkerboard solid. In contrast to the checkerboard case, the dipolar interactions beyond nearest-neighbors show a diagonal attraction along one lattice diagonal and repulsion along the other. These interactions give rise to a ground state of diagonal stripes with a period of $2\sqrt{2}a$, as shown in Fig. 2IIIc and d, with a corresponding peak in the structure factor at $\pm(\pi/2,-\pi/2)$ shown in Fig. 2IIIe. The observation of this diagonal stripe solid not only demonstrates the long-range dipolar nature of our interactions, but also highlights the low disorder of local chemical potential $\mu_i$ and the low temperature of our system compared to the maximum interaction energy of only $h \times \qty{10}{\Hz}$ with this dipole orientation.

\begin{figure*}
    \centering
    \includegraphics[width=\textwidth]{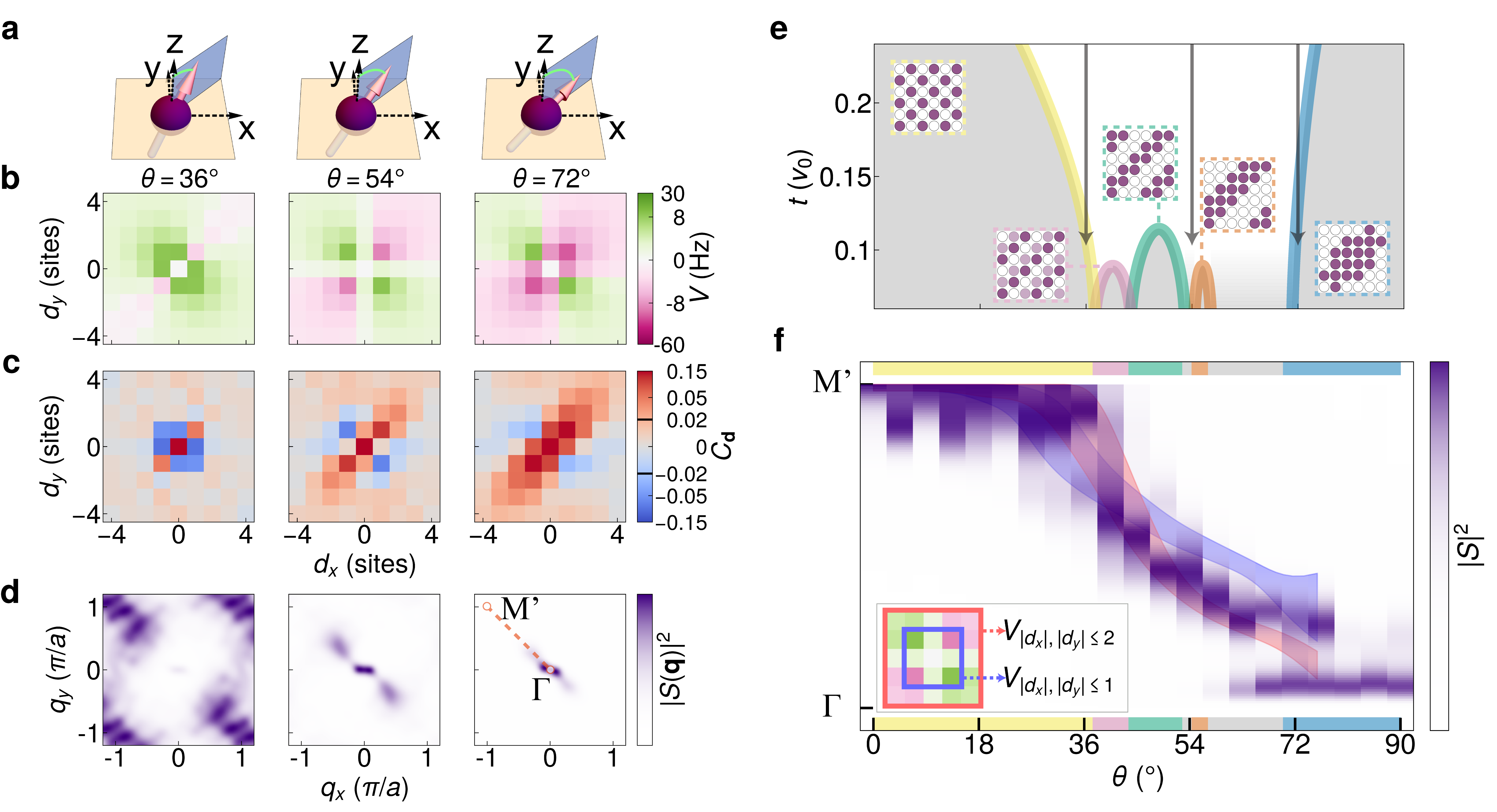}
    \caption{\textbf{Solids and global phase separation with spatial anisotropy.} For dipoles aligned diagonally with respect to the lattice vectors ($\phi$=\qty{45}{\degree}) different polar angles $\theta$ give rise to stripe solids with different periodicity. \textbf{a}, dipole orientation with the polar angle highlighted in green. \textbf{b}, anisotropic long-range dipolar interaction energy. \textbf{c}, measured connected density-density correlation. \textbf{d}, magnitude squared structure factors. \textbf{e}, Quantum Monte Carlo simulation of emerging phases. Gray shading indicates solids and white indicates a superfluid region. Each color border represents a different solid with the ordering shown in the inset. Between \qty{53}{\degree} and \qty{72}{\degree}, there are more solid phases without their boundaries marked out in color. Our adiabatic paths follow the gray arrows. \textbf{f}, Experimental structure factor along the orange diagonal line from marker $\Gamma$ to M' shown in the rightmost subfigure d, whose coordinates are $(q_x,-q_x)$. Shaded areas show the Quantum Monte Carlo simulation results with one standard deviation error. The blue area is a simulation only including nearest-neighbor and diagonal next-nearest-neighbor interaction terms ($|d_x|,|d_y|\leq1$). The red area includes beyond next-nearest-neighbor interaction terms ($|d_x|,|d_y|\leq2$). We find that the experimental data agrees better with the latter, indicating that long-range tails beyond next-nearest-neighbor terms play a significant role in our system.}
    \label{fig:figure3}
\end{figure*}

The example solid in Fig. 2III demonstrates that aligning the dipole orientation to an azimuthal angle $\phi=\qty{45}{\degree}$ between the x- and y-axis can result in isotropic nearest-neighbor interaction but anisotropic next-nearest-neighbor interaction, leading to translation symmetry breaking into a large unit cell \cite{Zhang2022}. Here, we further explore the ordering of the system at low temperature for different polar angles $\theta$ (Fig. 3a). We adiabatically reduce tunneling and approach the ground state with $t/V_0\approx0.1$ at different polar angles $\theta$ with the same $\phi=$\qty{45}{\degree} (Fig. 3a). As $\theta$ (green arc in Fig. 3a) increases, the observed diagonal stripe periodicity increases (Fig. 3c) and the structure factor peak (Fig. 3d) moves closer to the origin, indicating a larger unit cell. To gain theoretical understanding, we perform Quantum Monte Carlo simulations (see Methods) and identify the emerging phases for experimental parameters in Fig. 3e. 

To compare the observed structure factor with Quantum Monte Carlo simulations, we plot the measured density structure factors $S(\mathbf{q})$ along the straight line from $\mathrm{\Gamma}$ to $\mathrm{M}'$ point in the Brillouin zone (the dotted diagonal line shown in the rightmost subfigure of Fig. 3d), whose coordinates are $(q_x,-q_x)$, and plot the results in Fig. 3f in increments of \qty{4.5}{\degree}. The structure factor indicates checkerboard order between $\theta=\qty{0}{\degree}$ and $\theta\approx\qty{30}{\degree}$. Then, the peak location of the density structure factor gradually moves towards the origin until $\theta\approx\qty{80}{\degree}$. At this point, the attractive long-range interaction dominates over the repulsive one. We observe that the ground state of our finite system is no longer a diagonal stripe solid, but instead, it is a self-organized state with the lattice separated into unity-filled and empty regions, each occupying only half of the sites in the region of interest. We denote this a global phase separation state, where the unity filled region is a self-bound insulator. This state is fundamentally distinct from a unity-filled Mott insulator, as the system would form a half-filling superfluid in the absence of dipolar interactions.

The various types of diagonal stripes that can form between the checkerboard solid and the global phase separation state exemplify the long range of the dipolar interactions beyond next-nearest-neighbors. We simulate the dipolar interaction cutoff with $|d_x|,|d_y|\leq1$ (blue shade, 3 by 3 box, includes interactions only for nearest-neighbor and next-nearest-neighbor) and $|d_x|,|d_y|\leq2$ (red shade, 5 by 5 box, includes interactions beyond next-nearest-neighbor). The tail of the long range interactions plays an important role in the self-organization of stripe solids with long periodicity when $\theta$ is between $\qty{50}{\degree}$ and $\qty{70}{\degree}$, causing the significant difference between the simulation results with different cutoff ranges. Remarkably, we observe our data to agree with the simulation including beyond-next-nearest-neighbor interactions (red shade), highlighting the important role of long-range interaction terms.

\begin{figure*}
    \centering
    \includegraphics[width=\textwidth]{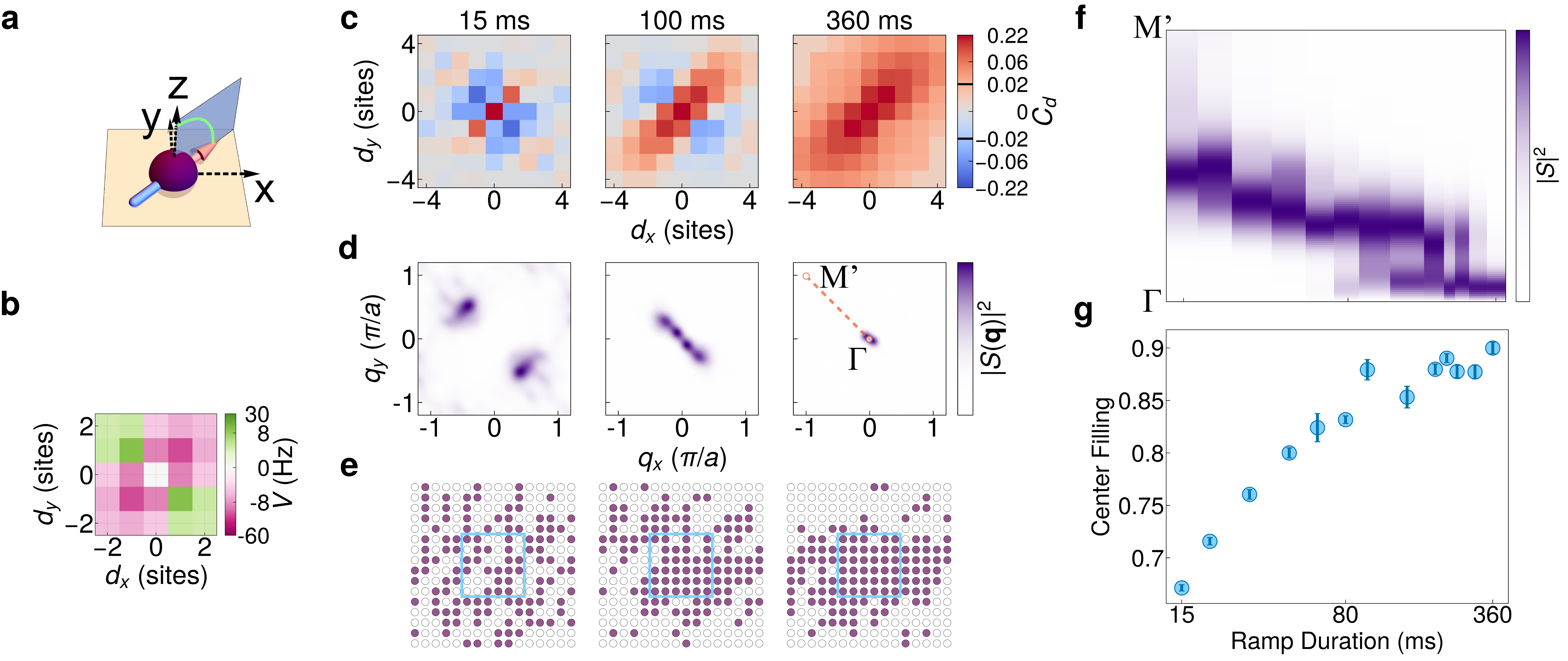}
    \caption{\textbf{Out-of-equilibrium dynamics.} After ramping non-adiabatically into the global phase separation state, we observe diagonal stripe solids with different periodicity depending on the ramp speed, which demonstrates the rich manifold of metastable states lying above the global phase separation state. The dipole is oriented diagonally in the $x$-$y$ -plane  at $\phi$= \qty{45}{\degree} and $\theta$= \qty{90}{\degree} (\textbf{a}), leading to the anisotropic dipolar interaction energy (\textbf{b}). The connected density-density correlation (\textbf{c}), the structure factor (\textbf{d}), and single-shot images (\textbf{e}) show the varying periodicity for different ramp durations of \qty{15}{\ms}, \qty{100}{\ms}, and \qty{360}{\ms}. We plot the diagonal structure factor (\textbf{f}) and the center (within the blue box in subfigure e) filling ratio (\textbf{g}, error bar denotes the standard error of the mean) to show the emerging phase separation as the filling at the center approaches unity. }
    \label{fig:figure4}
\end{figure*}

\textit{Dynamic ramp into the metastable states above global phase separation.} We explore the rich out-of-equilibrium physics that emerge from the dipolar interactions (Fig. 4). In particular, the global phase separation state when $(\theta,\phi)=(\qty{90}{\degree},\qty{45}{\degree})$ exhibits a variety of low-lying metastable states, which is a characteristic of long-range systems \cite{Defenu2021a, Trefzger2008}. To explore these metastable states, we dynamically ramp the system starting from a half-filling superfluid using a linear ramp of lattice depth, resulting in an exponential ramp in tunneling energy from $h \times \qty{300}{\Hz}$ to $h \times \qty{3}{\Hz}$. By varying the ramp time, we observe diagonal stripes with different periodicities (Fig. 4c). When the ramp is faster than \qty{100}{\ms}, the structure factor exhibits the signature of diagonal stripes, and the atoms are sparsely arranged throughout the analysis region (Fig. 4e left).  These diagonal stripes are among the lowest-lying metastable states because the dipolar interaction (Fig. 4b) strongly biases diagonal stripes aligned with the dipole direction. As the ramp becomes more adiabatic, the structure factor peaks migrate towards the origin (Fig. 4f) until they become limited by the system size, and the atoms form a state that is close to unity-filled (Fig. 4e right). Due to finite temperature, we still observe sporadic excitations from the ground state in the form of a missing atom in the center of the cloud. These excitations do not exhibit significant spatial structure, unlike the metastable diagonal stripe patterns. Overall, the long-range nature of the dipolar interaction is responsible for the rich out-of-equilibrium physics observed in this system.

\textit{Conclusion and Outlook.} We observe various dipolar quantum solids and global phase separation using ultracold atoms with single-site resolution, demonstrating the rich phases caused by the anisotropic long-range dipolar interactions. We show evidence of metastable states, which we access by changing the ramp speed across the phase transition from superfluid to the global phase separation state. These observations mark the beginning of studying itinerant, strongly correlated quantum phases originating from dipolar interactions. The dipolar quantum gas microscope provides a flexible platform to examine a host of quantum phase transitions between different states in the dipolar system and dynamics across such transitions, such as first-order phase transitions and transitions exhibiting intermediate ``microemulsion'' phases \cite{Spivak2004, Zhang2015}. Furthermore, metastable states represent an exciting frontier in many-body quantum systems and open the gate to quantum orders not stable in equilibrium \cite{Sahay2022}. Moreover, going beyond the hard-core boson limit and allowing multiple particles to occupy the same site promotes supersolid phases to occupy a much larger region in the phase diagram \cite{Iskin2011, Grimmer2014}, and gives rise to Haldane insulators in one dimension \cite{DallaTorre2006}. Leveraging the anisotropic long-range interactions in this system will allow examination of spin liquids \cite{Yao2018} and fractionalization \cite{Prem2018, Mao2022}. Finally, employing fermionic species of magnetic atoms will enable the realization of extended Fermi-Hubbard models to study spinful itinerant systems with dipolar interactions, which can show a wide variety of phenomena including bond order waves \cite{Julia2022} and ultralong-range order \cite{vanLoon2015}.

\nocite{*}

\section*{Methods}

\subsection*{Model calibration}
We study the extended Bose-Hubbard model in the hard-core boson limit as the on-site interaction energy $U$ in our system is much larger than tunneling energy $t$, dipolar energy scale $V_0$ and chemical potential $\mu_i$. Thus, we neglect the on-site interaction term $\frac{U}{2}\sum_i\hat{n}_i(\hat{n}_i-1)$ in the standard Bose-Hubbard Hamiltonian. We estimate the on-site interaction energy $U$ to be $h \times \SIrange{1}{2}{\kHz}$, depending on the atomic dipole orientation. We measure $U$ by modulating the lattice intensity and observing atom loss in the lattice when the modulation frequency is close to $U$ \cite{Baier2016}. We also compute $U$ by taking into account the vertical Wannier function width as well as the dipole orientation and find good agreement with previous measurements \cite{Patscheider2022}. Such high $U$ results in a negligible super-exchange energy of $h \times \qty{0.01}{\Hz}$ ($4\times 10^{-4} V_0$) at typical tunneling energies at which we examine Dipolar Quantum Solids. To calibrate tunneling $t$ for the data in this paper, we measure the lattice depth by lattice modulation \cite{Baier2016}. From the measured lattice depth, we numerically obtain the lattice band structure and estimate tunneling from the ground band width \cite{Bloch2008}. We compute the density-induced tunneling \cite{Dutta2015} to be an order of magnitude smaller than the single-particle tunneling $t$, so we ignore the term in the Hamiltonian. The variation of the chemical potential $\mu_i$ in our system consists of the global harmonic confinement introduced by the lattice beams and the site-to-site disorder in the center of the analysis region. We measure the global harmonic confinement by measuring the oscillation frequency of the atomic cloud after releasing the atoms from a tight dipole trap. We estimate the chemical potential disorder in the analysis region of interest (around 100 sites) by comparing measured lattice filling to exact diagonalization simulation. Specifically, we ramp the lattice to different lattice depths and measure the standard deviation of the average atom number filling.  We compare this measurement with a simulation of a $4\times4$ site system by exact diagonalization, assuming the disorder follows a uniform distribution \cite{Zhang2018}. We estimate the chemical potential $\mu_i$ disorder to be $h \times \qty{3}{\Hz}$ from site to site, in addition to a global harmonic confinement. We compute the dipolar interaction strength $V_{i,j}$ at different distances, by taking into account the non-zero width of the Wannier function, which can modify the interactions between the nearest-neighbor sites significantly \cite{Korbmacher2023}.

\subsection*{Lattice stability}

At several points in the experimental sequence, we transfer atoms between different sets of lattices, which requires stable relative lattice phase for reliable operation. To achieve high stability, we use the vacuum chamber as a reference for all the lattices in our experiment. The 2D small-spacing \qty{532}{nm}-wavelength lattices are retro-reflected from mirrors attached to the vacuum chamber. The 2D tunable-spacing \qty{488}{nm}-wavelength accordion lattices are sent through the objective, which is directly mounted inside the vacuum chamber. The vertical \qty{1064}{nm}-wavelength lattice is retro-reflected from a mirror mounted on top of the in-vacuum objective. To match the phase of the 2D small-spacing retro-reflected lattices to that of the 2D accordion lattices, we use a thin AR coated window mounted on a galvanometer to tune the path length difference of the accordion beams. Once the system is warmed up for two hours, the phases of the lattices stay stable for more than 24 hours.

\subsection*{Reducing local chemical potential disorder}

Achieving chemical potential disorder smaller than the dipolar interaction energy scale $V_0$ means that the intensity of scattered light interfering with the main lattice beam has to be less than 1 part per million --- an extraordinarily low ratio to achieve experimentally. We find the main source of chemical potential disorder in our experiment to be the scattered vertical lattice light. Therefore, we tune the intensity of the vertical lattice to be as low as possible, while still keeping the atoms in a single layer of the vertical lattice during the experimental sequence \cite{Aeppli2022}. The energy difference due to gravity in adjacent layers of the vertical lattice prevents resonant tunneling of atoms. Experimentally, we observe that a vertical lattice depth of around 7 $\Er$ maximizes the dipolar quantum solid order, where $\Er$ is the vertical lattice recoil energy. The lifetime of the atoms at such lattice depth is on the order of a second \cite{Gluck1999}. Thus, our experimental lattice ramps of a few hundred milliseconds are short enough to keep most of the atoms in the lattice.

\begin{figure}
    \centering
    \includegraphics[width=0.45\textwidth]{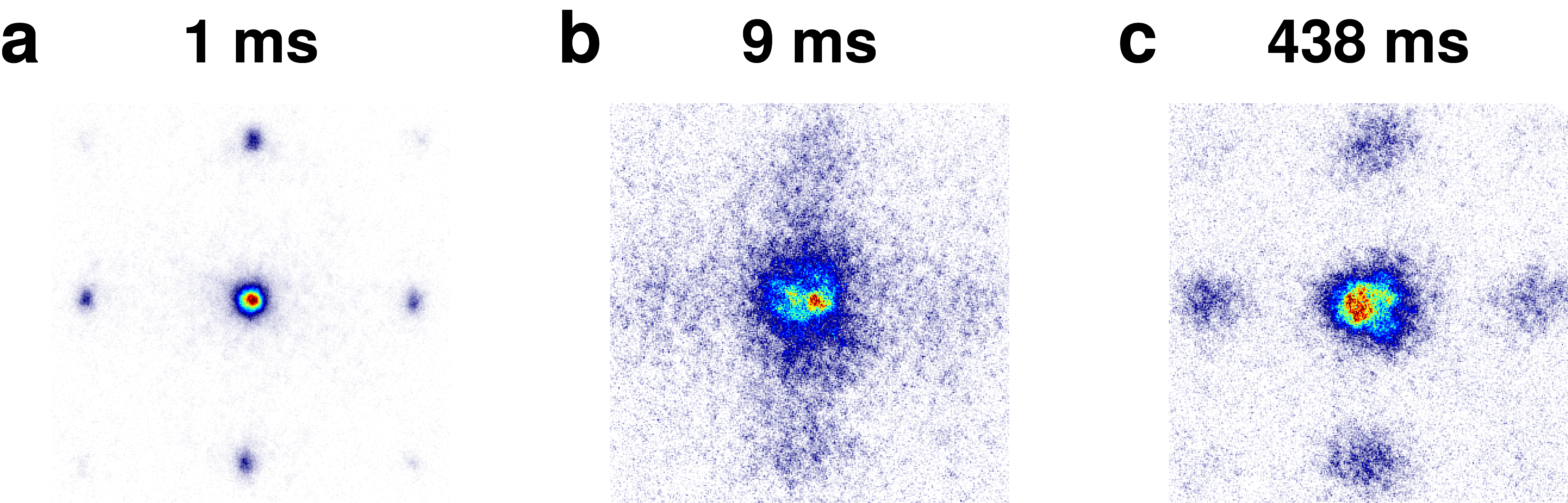}
    \caption{\textbf{Adiabaticity of the lattice ramp.} We probe the adiabaticity of the lattice ramp by varying the duration as we follow the solid and dashed arrow to return to the star position at 7 $\Er$ in Fig. 1d. When the ramp duration is very short at \qty{1}{\ms} (\textbf{a}), we see sharp coherence peaks in the time-of-flight image. As we slow down the ramp to a duration of \qty{9}{\ms} (\textbf{b}), the coherence peaks in the time-of-flight image are the least resolved. Further increasing the ramp duration up to \qty{438}{\ms} (\textbf{c}), we observe well-resolved coherence peaks again. The peaks are less sharp compared to those in \textbf{a}, possibly due to the decoherence and atom loss during the ramps, which in total takes almost \qty{900}{\ms}. These averaged images demonstrate that, with the ramp duration on the order of \qty{100}{\ms} in this paper, the system is in the adiabatic regime.}
    \label{fig:figure7}
\end{figure}

\subsection*{State preparation}

We first create a Bose Einstein Condensate in less than a second as described in \cite{Phelps2020}. The single-chamber experiment design removes the complexity of transport. Next, we compress the atoms into a thin sheet using a vertical \qty{532}{nm}-wavelength accordion lattice that goes through an aspheric lens mounted inside the vacuum chamber in \qty{400}{\ms}. Then we transfer the atoms to a single layer of the vertical \qty{1064}{nm}-wavelength retro-reflected lattice. We control the final atom number in the lattice by tuning the power of an additional optical dipole trap as we turn on a magnetic field gradient to pull the rest of the atoms out of the trap. This process takes \qty{600}{\ms}. We observe a few percent of atom number fluctuations (standard deviation) from shot to shot and do not post-select for exact half-filling (except Fig. 7).  After turning off the field gradient, we rotate the bias magnetic field to the desired orientation. Finally, we ramp up the 2D small-spacing retro-reflected lattice exponentially for \qty{100}{\ms} to 8 $\Er$, and then linearly for a few hundred milliseconds to 27 $\Er$. To demonstrate adiabaticity of the lattice ramp, we probe the coherence peaks as we change the duration of the linear ramp (Fig. 5).

\subsection*{Imaging procedure}
At the end of each experiment, we freeze the dynamics in the small-spacing optical lattice by ramping up the lattice power so that the tunneling changes from roughly $h \times \qty{3}{\Hz}$ to $h \times \qty{0.3}{\Hz}$ in $\qty{100}{\us}$. We then transfer the atoms from the small-spacing lattice to the 2D accordion lattice in $\qty{6}{\ms}$. Next, we expand the 2D accordion lattice spacing from \qty{266}{\nm} to \qty{3}{\um} in $\qty{80}{\ms}$ \cite{Li2008}. We perform fluorescence imaging of the atoms by exposing them to highly-saturated \qty{401}{\nm} resonant beams for \qty{8}{\us} without cooling or trapping \cite{Bergschneider2018}. Thanks to the large \qty{30}{\MHz} linewidth of the imaging transition, we detect roughly 50 photoelectrons during the exposure time using an electron-multiplying CCD camera. To minimize the net momentum on the atoms exerted by the imaging beams, we expose the atoms to two counter-propagating imaging beams. 
We alternately pulse the two counter-propagating beams to eliminate the effect of a standing wave created when both beams are applied at the same time \cite{Bergschneider2018}. We keep the beam intensity much higher than the saturation intensity $I_{\mathrm{sat}}$, such that the scattering rate and thus the average momentum imparted by each beam onto the atoms is stable without the need for closed-loop feedback of the beam power. During imaging, the atoms experience movement due to stochastic momentum kicks imparted by the spontaneously emitted photons; however, the random walk moves the atoms by less than half the accordion lattice spacing in most cases. 

\begin{figure}
    \centering
    \includegraphics[width=0.45\textwidth]{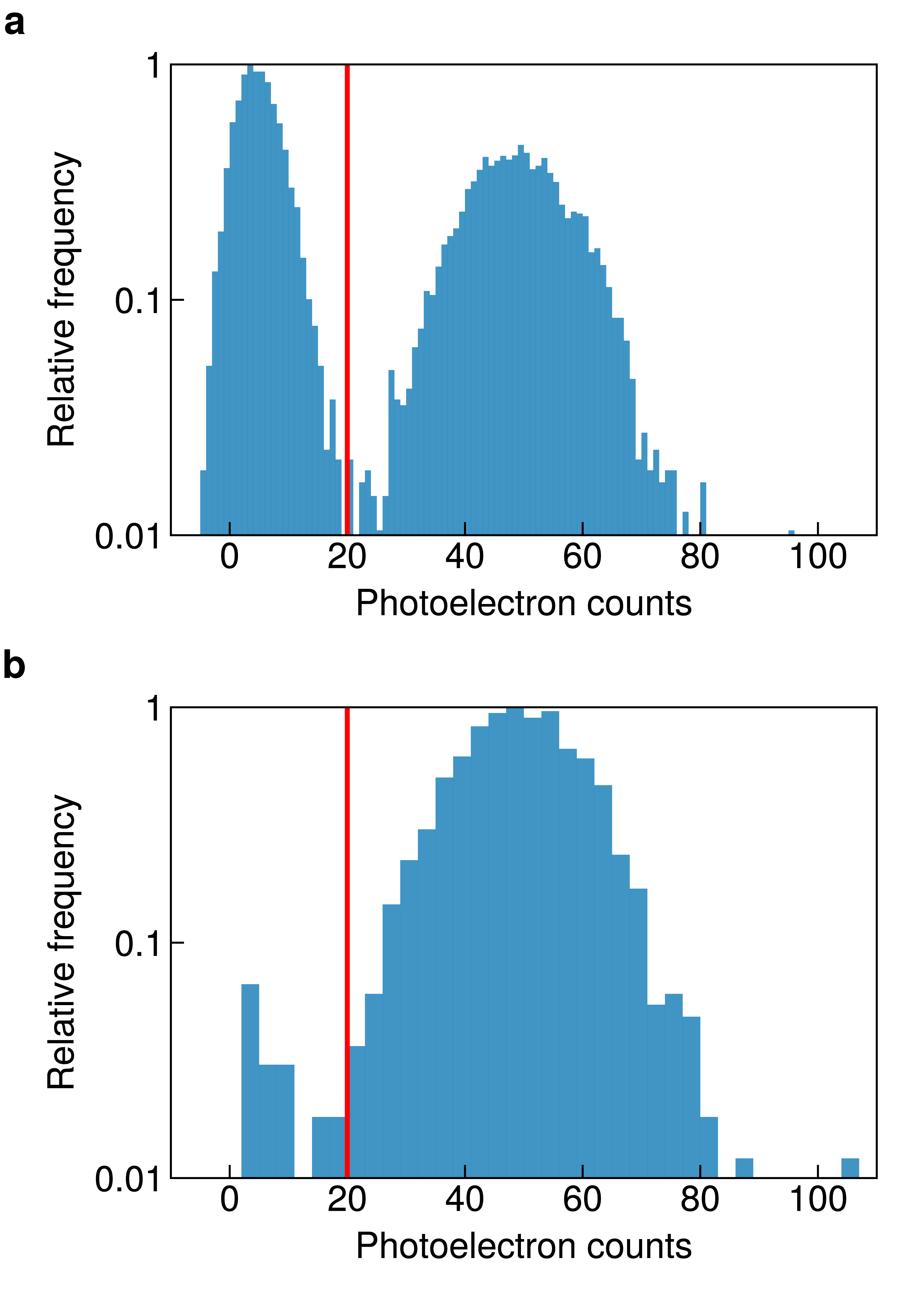}
    \caption{\textbf{Histogram for digitization of occupation number.} We perform high-fidelity site-resolved imaging after expanding the 2D accordion lattice to \qty{3}{\um} spacing. \textbf{a}, half-filling histogram. The fidelity to distinguish between 0 and 1 filling per site after expanding the 2D accordion lattice is more than \qty{99}{\percent}. \textbf{b}, unity-filling histogram. The efficiency of transferring atoms to the 2D accordion lattice and expanding is more than \qty{98}{\percent}. 
    }
    \label{fig:figure5}
\end{figure}

\subsection*{Imaging fidelity}

To distinguish between one and zero atoms on each lattice site, we sum up the signals of all camera pixels corresponding to the site and then digitize the number of atoms per site with a precalibrated threshold of 20 photoelectrons per site. In Fig. 6a, we show the histogram of total photoelectron counts of the same site in 9000 shots collected over 10 hours. The two Gaussian-like peaks are well separated and correspond to a fidelity to distinguish between 0 and 1 atoms above \qty{99}{\percent}. Characterizing loss during the imaging protocol, including the loss during transfer into the blue accordion lattice as well as during the expansion of the blue accordion lattice, is more challenging. We prepare a Mott insulator with one particle per site ($n=1$) and estimate an upper bound of the loss if we assume that the Mott insulator we made has \qty{100}{\percent} fidelity. In Fig. 6b, we load an $n=1$ Mott insulator and show the histogram of total photoelectron counts within all lattice sites in the center 10 by 10 sites. We detect the filling of the Mott insulator to be higher than \qty{98}{\percent}, giving us an upper bound of loss to be \qty{2}{\percent}.

\subsection*{Numerical simulations}
We perform worm-type Quantum Monte Carlo simulations based on an existing package \cite{Sadoune2022} that we modify to include long-range dipolar interactions. We benchmark our simulations with finite temperature exact diagonalization \cite{Weinberg2016} for a $4\times4$ site system. For typical experimental parameters used in this work, we find that the auto-correlation time is short after thermalization: about 10 measurements for the density-density correlations compared to the 5000 measurements we perform. The short auto-correlation time is partially due to the metastability of states in the long-range interacting system's Hilbert space. To avoid this, we thermalize 64 different seeds and average the results. When estimating the temperature of the solids, we set the boundary condition to open and include measured harmonic confinement in the simulation, but do not include local chemical potential disorder. For the phase diagram in Fig. 3e, we set the temperature to $T=0.2V_0/k_B$ (\qty{0.3}{nK}), the system size to 20 by 20 sites, and the boundary condition to open. In addition, we assume flat chemical potential with no harmonic confinement and no disorder. For all of the phases, we use the location of the peak of the structure factor as the order parameter, which we then compare with the location of the structure factor peak in the classical limit of no tunneling. The simulation of the structure factor in Fig. 3f has the same conditions as the phase diagram, but uses the experimental uncertainty in tunneling to determine error bars. Simulations at lower temperatures do not shift the location of the structure factor peak, and only serve to narrow it, allowing us to simulate the emerging phases down to $t/V_0=0.05$ for Fig. 3e.

\subsection*{Data analysis}
All density correlation and structure factor data shown in this paper are computed from a square box analysis region, whose size ranges from 11 by 11 sites to 15 by 15 sites, resulting in a full $C_\mathbf{d}$ matrix of 23 by 23 sites to 31 by 31 sites depending on the particular data set. We only show the $C_\mathbf{d}$ in the figures within $|d_x|,|d_y|\leq4$ sites because most of the ordering we study decays within 4 sites. We compute the connected density-density correlation instead of the disconnected one because we want to distinguish the classical solid with a static phase from the solid with an uncertain phase. If static chemical potential disorder were to pin the solids down to one of the static phase, we would no longer see signal in the connected density-density correlation. The observed correlation decays exponentially over distance, so we mainly use a logarithmic color scale and connect the positive and negative logarithmic scales with a linear scale. We compute the structure factor from the connected density-density correlation matrix extracted from typically up to $|d_x|,|d_y|\leq9$ sites. To obtain the diagonal structure factors shown in Fig. 3f and Fig. 4f, we integrate the structure factor perpendicular to the straight line (shown in Fig. 3d and Fig. 4d) with a Gaussian envelope centered on the straight line. The center filling is calculated by finding the maximum filling ratio in a 6 by 6 box among shots with certain filling range (40 to 45 per cent) in the whole analysis region of 16 by 16 sites.

\subsection*{Solid temperature}
We compare the density-density correlation measured in experiment with the Quantum Monte Carlo simulation to estimate the temperature of our system. For the stripe solid, we estimate the temperature to be \qty{0.8}{nK}; for the checkerboard solid, we estimate the temperature to be \qty{0.7}{nK}. We compute the critical temperature to be around \qty{0.5}{nK} for the stripe solid and estimate the critical temperature to be lower for other types of solids examined in this work. In order to fix the particle density we sweep both the temperature and chemical potential and then for each temperature use the chemical potential that replicates the experiment's filling fraction in the central 16 sites. All solids in this paper exhibit connected density-density correlation that decay exponentially as we increase $d$, indicating that the system is above the critical temperature of the solids or that chemical potential disorder and harmonic confinement is too large \cite{Zhang2018}. The exponential fits to the density correlations ($d>1$) give the correlation length. Specifically, the stripe density-density correlation decays exponentially with a correlation length of $\num{2.22 +- 0.05}$ sites for $\mathbf{d}=(d_x,0)$ and $\num{1.21 +- 0.15}$ sites for $\mathbf{d}=(0, d_y)$. The checkerboard correlation decays exponentially with a correlation length of $\num{0.82 +- 0.05}$ sites isotropically. The diagonal stripe correlation decays with a correlation length of $\num{0.6 +- 0.2}$ along the diagonal directions. The observed correlations decay exponentially in the experiment, but they are the same ones that are established at long range in the solid phases, and they can be observed in a whole sub-region in some single-shot images (Fig. 2c).

\subsection*{Solid lifetime}
After we ramp into the solid phases, we hold the atoms for varying duration and measure the overlap of the connected density-density correlation with the perfect solid correlation. We observe the overlap to decay roughly exponentially with a lifetime of \qty{1}{\second} for the stripe solid when the atomic dipole points along one lattice direction and \qty{0.7}{\second} for the checkerboard solid. The finite lifetime of our Wannier-Stark state in the vertical lattice as well as heating from technical sources like laser noise and scattering can contribute to the decay of the solid order. Such a lifetime enables future work exploring the phase transitions between dipolar quantum solids and supersolids, since such experiments require adiabatic ramps of the dipole orientation when tunneling energy is small.

\begin{figure}
    \centering
    \includegraphics[width=0.45\textwidth]{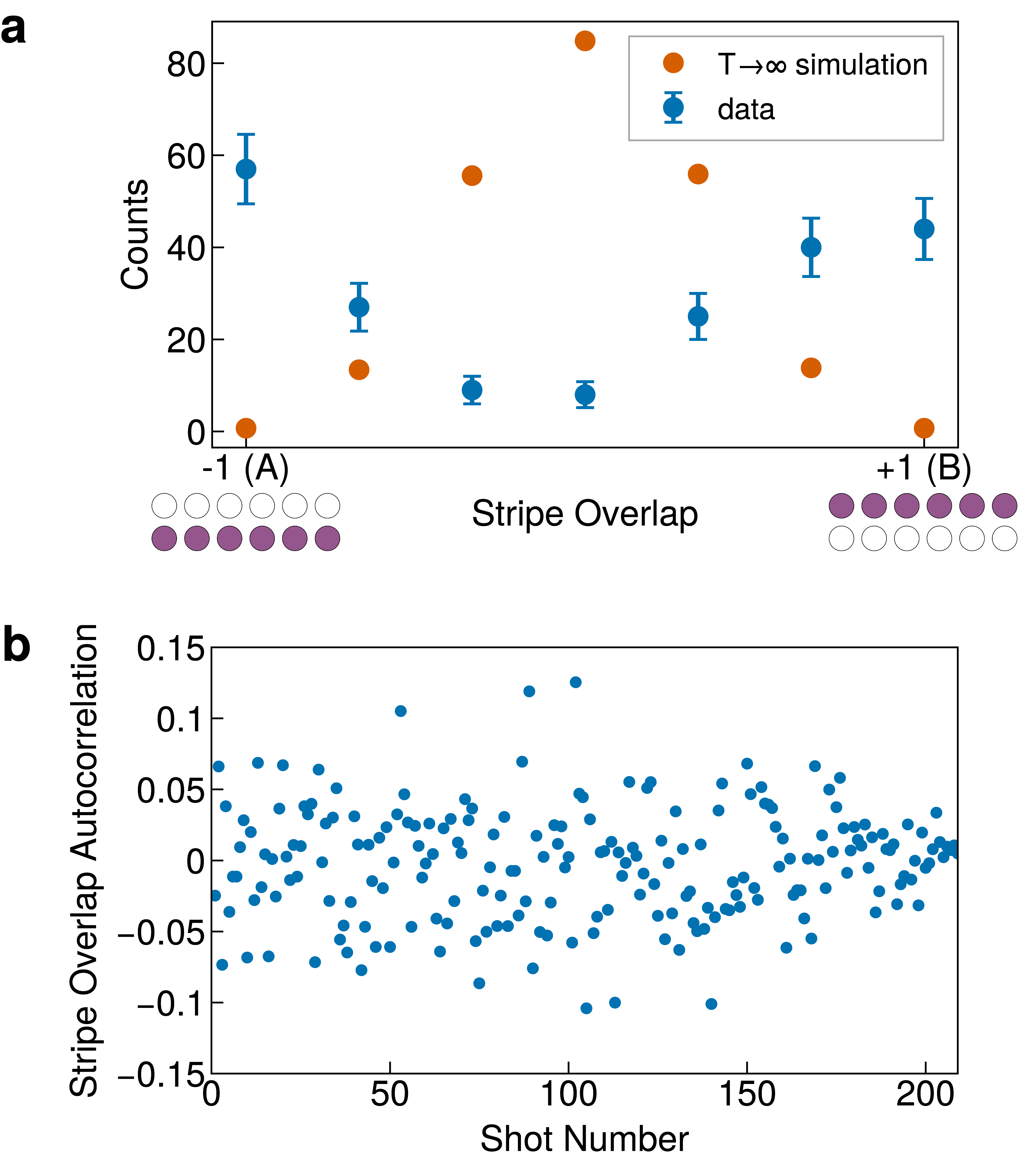}
    \caption{\textbf{Stripe overlap in a 2 by 6 box.} We demonstrate the bimodal distribution of the macrostate that is temporally uncorrelated. \textbf{a}, histogram of the overlap of the stripe order with the single shot data (blue) and simulation of the infinite temperature state (orange). \textbf{b}, Auto-correlation of the stripe overlap data. 
    }
    \label{fig:figure6}
\end{figure}

\subsection*{Spontaneous symmetry breaking of stripe solids}

With the site-resolved single-shot images, we can examine the spontaneous symmetry breaking as we transition into a stripe solid. We choose a 2 by 6 site region in our system and study the stripe ordering after post-selection on having exactly 6 atoms over these 12 sites. We define a value that measures the distance of a single shot image to the stripes of phase A and B (bottom of Fig. 7a). With the simulated infinite temperature state, we see most of the values to be centered in between the two stripe patterns A and B, as the orange points show in Fig. 7a. But in our experiment with the stripe solid, we observe the perfect stripe ordering more frequently, leading to a bimodal distribution. To demonstrate that there is no temporal correlation between the two stripe patterns and support the claim that the symmetry breaking is random, we calculate the auto-correlation of the stripe overlap values over different shots, and demonstrate that the overlap value is stochastic and shows no auto-correlation peaks (Fig. 7b). The above provides evidence that our stripe solid is a spontaneous symmetry breaking state.

\bibliography{DipolarQuantumSolidsMain}% Produces the bibliography via BibTeX.

\section*{Data availability}
The data that support the findings of this study are available from the corresponding author on reasonable request.

\section*{Acknowledgments}
We wish to acknowledge Vassilios Kaxiras, Anant Kale, Muqing Xu, Maximilian Sohmen, Manfred Mark, and Yicheng Bao for help on building the experiment. We wish to acknowledge Rahul Sahay, Barbara Capogrosso-Sansone, Eun-Ah Kim, Lukas Homeier, Annabelle Bohrdt, Fabian Grusdt, Martin Lebrat, Tilman Esslinger, Matjaz Kebric, Subir Sachdev, and Carlos A. R. Sa de Melo for helpful discussions. We are supported by U.S. Department of Energy Quantum Systems Accelerator DE-AC02-05CH11231, National Science Foundation Center for Ultracold Atoms PHY-1734011, Army Research Office Defense University Research Instrumentation Program W911NF2010104, Office of Naval Research Vannevar Bush Faculty Fellowship N00014-18-1-2863, and Defense Advanced Research Projects Agency Optimization with Noisy Intermediate-Scale Quantum devices W911NF-20-1-0021. A.D. acknowledges support from the NSF Graduate Research Fellowship Program (grant DGE2140743). The computations in this paper were run on the FASRC Cannon cluster supported by the FAS Division of Science Research Computing Group at Harvard University.

\section*{Author contributions}
L.S., A.D., M.S, R.G., S.F.O., A.K, A.H.H., G.A.P., S.E., S.D. and O.M. contributed to building the experiment set-up. L.S., A.D., M.S. and O.M. performed the measurements and analyzed the data. A.D. performed theoretical analysis. M.G. conceived the experiment in collaboration with F.F. M.G. supervised all works. All authors discussed the results. L.S., A.D., M.S., R.G., S.F.O., F.F., O.M. and M.G. contributed to the manuscript.

\section*{Competing interests}
M.G. is a cofounder and shareholder of QuEra Computing. All other authors declare no competing interests.

\end{document}